\documentclass[10pt,a4]{article}
\usepackage{times,fancyhdr}
\usepackage[dvips]{graphicx}
\usepackage{amsmath}
\usepackage{amssymb}
\usepackage{array}
\usepackage[small,bf]{caption}
\usepackage{color}
\usepackage[yyyymmdd,hhmmss]{datetime}
\usepackage{fancyhdr}
\usepackage{fancyvrb}
\usepackage{hyperref}
\usepackage{longtable}
\usepackage{setspace}
\usepackage{framed}
\usepackage{titlesec}
\usepackage{titling}
\usepackage{xcolor}
\usepackage{authblk}
\usepackage{booktabs}
\usepackage{ragged2e}
\usepackage{lineno} 
\usepackage{natbib}
\usepackage{comment}
\let\cite\citep

\fancyheadoffset[R]{0pt}
\fancyfootoffset[R]{0pt}
\definecolor{orange}{rgb}{1.0,0.5,0.0}
\definecolor{aqgr}  {rgb}{0.0,1.0,0.6} 
\definecolor{viol}  {rgb}{0.8,0.6,0.8}
\definecolor{figdr} {rgb}{1.0,1.0,1.0} 
\definecolor{colne} {rgb}{1.0,0.0,1.0} 
\definecolor{coldr} {rgb}{1.0,0.8,0.0} 
\definecolor{colop} {rgb}{0.5,1.0,1.0} 
\definecolor{colok} {rgb}{0.7,1.0,0.7} 
\definecolor{colcg}{rgb}{0.0,0.0,0.0}

\newcolumntype{L}[1]{>{\raggedright\hspace{0pt}}p{#1}}
\newcolumntype{P}[1]{>{\centering\arraybackslash}p{#1}}
\newcommand\parop[1]{\colorbox{colop}{\textbf{#1}}}
\newcommand\parok[1]{\colorbox{colok}{\textsc{#1}}}
\newcommand\neudef[1]{\textcolor{black}{\textit{#1}}}
\def\afhead{0}

\title{\vspace{0.0cm}\bfseries{\textsc{ 
Evolution, the mother \\ 
of age-related diseases
}}}

\author{}  
\vspace{-1.0cm} \date{}
\begin{document}

\pretitle{%
\begin{center}\LARGE
\vskip -2.2cm
\rule{\textwidth}{2.0pt}
\par
\vskip 0.5cm
}
\posttitle{
\par
\rule{\textwidth}{2.0pt}
\end{center}
\vskip -0.0cm
}
\maketitle

\vspace*{-2.5cm}
\begin{center}
\begin{tabular}{P{5.0cm} P{0.0cm} P{0.0cm}}
\textbf{Alessandro Fontana} & & \\
\texttt{fontalex00@gmail.com} & & 
\end{tabular}
\end{center}
   
\vspace*{0.0cm}
\begin{abstract}
\normalsize
\if\afhead2 {\parop{xxxx}} \fi
The evolutionary origins of ageing and age-associated diseases continue to pose a fundamental question in biology. This work is concerned with a recently proposed framework, which conceptualises development and ageing as a continuous process, driven by genetically encoded epigenetic changes in target sets of cells. According to the Evolvable Soma Theory of Ageing (ESTA), ageing reflects the cumulative manifestation of epigenetic changes that are predominantly expressed during the post-reproductive phase. These late-acting modifications are not yet evolutionarily optimised but are instead subject to ongoing selection, functioning as somatic ``experiments'' through which evolution explores novel phenotypic variation. These experiments are often detrimental, leading to progressive physical decline and eventual death, while a small subset may produce beneficial adaptations, that evolution can exploit to shape future developmental trajectories. 
According to ESTA, ageing can be understood as evolution in action, yet old age is also the strongest risk factor for major diseases such as cardiovascular disease, cancer, dementia, and metabolic syndrome. We argue that this association is not merely correlational but causal: the same epigenetic process that drive development and ageing also underlie age-associated diseases. Growing evidence points to epigenetic regulation as a central factor in these pathologies, since no consistent patterns of genetic mutations have been identified, whereas widespread regulatory and epigenetic disruptions are observed. From this perspective, evolution is not only the driver of ageing but also the ultimate source of the diseases that accompany it, making it the root cause of most age-related pathologies.
\end{abstract}


\section{Introduction}

\if\afhead1 {\parok{mistery: diseases}} \fi 
Despite decades of research, major diseases such as cardiovascular disease, cancer, neurodegenerative disorders and diabetes remain profound mysteries in biology and medicine. While various risk factors have been identified, including genetic predispositions, lifestyle choices, and environmental influences, their ultimate causes remain elusive. A striking commonality among these conditions is their strong association with ageing: as lifespan increases, so does disease prevalence. Together, these age-related diseases account for the vast majority of global morbidity and mortality, with estimates suggesting that they are responsible for over 70\% of deaths worldwide, primarily affecting older adults \cite{WHO2023NCD}.

\if\afhead1 {\parok{mistery: ageing}} \fi 
Ageing itself is one of the greatest biological open problems, with the relative contributions of genetic and non-genetic factors still largely unresolved. \textit{Stochastic theories} \cite{Lopezotin13} propose that ageing arises from cumulative damage caused by environmental stressors or metabolic byproducts, such as free radicals, akin to the gradual deterioration of mechanical systems. In contrast, \textit{programmed theories} \cite{Blagosklonny06} suggest that ageing follows a genetically encoded biological schedule, though its progression may be modulated by external factors. \textit{Epigenetic ageing theories} \cite{Horvath13} posit that age-related decline is primarily driven by shifts in gene expression patterns rather than direct changes to the DNA sequence, leading to cellular dysfunction and tissue degeneration.

\if\afhead1 {\parok{mistery: evolution}} \fi 
Evolution is yet another biological puzzle, as it seeks to explain the origin and diversification of life through genetic variation and natural selection. However, significant gaps in the evolutionary record, particularly regarding the sudden appearance of complex biological structures, have led some researchers to explore alternative explanations. In \cite{Klinghoffer15}, scholars critically examine these challenges, questioning whether traditional evolutionary mechanisms alone can account for the intricate information encoded in DNA and the rapid emergence of novel body plans. Proponents of Intelligent Design \cite{Dembski98} argue that the complexity and specificity observed in biological systems suggest the presence of an underlying intelligence guiding these processes.

\if\afhead1 {\parok{mistery: ageing and evolution}} \fi 
If evolution and ageing are mysteries in their own right, their intricate relationship is even more perplexing. The idea that ageing is a nonadaptive process (a byproduct rather than a trait shaped by natural selection) is strongly supported within the field \cite{Vijg16essence}. The prevailing view suggests that ageing persists because evolutionary pressure diminishes with age, making ageing a byproduct of limited investment in cellular repair and maintenance in favour of early-life evolutionary benefits, or an effect of trade-offs involving genes with pleiotropic effects. This is the concept behind the current state-of-the-art evolutionary ageing theories \cite{Medawar52, Williams57, Kirkwood77}. 

\if\afhead1 {\parok{model}} \fi 
In this work, we will try to address all these questions, through a novel hypothesis on the relationship between ageing and evolution. Our approach builds on the \neudef{Epigenetic Tracking (ET)} model \cite{Font08}, a framework that enables the interpretation of various biological phenomena \cite{Font12b, Font23a, Font14}. A key feature of ET is its unified approach to development and ageing, both governed by genetically encoded events triggered by time-dependent timers that evolution can modify by advancing or delaying their activation. While our hypothesis is grounded in the ET model, its general principles can be extended to any framework that incorporates these core features

\if\afhead1 {\parok{structure}} \fi 
This paper is organised into six sections. After this introduction, Section 2 introduces the ET model, explaining its key principles and implications. Section 3 outlines the hypothesis on ageing and evolution which constitutes the backbone of this work. Section 4 presents an overview of major human diseases, highlighting their association to the ageing phenomenon. Section 5 discusses the link between evolution and age-related diseases. Finally, Section 6 presents the conclusions and outlines potential directions for future research.

\section{The ET model of development}
\label{sec:model}

\begin{figure*}[t] \begin{center} \hspace*{-0.2cm}
\includegraphics[width=17.00cm]{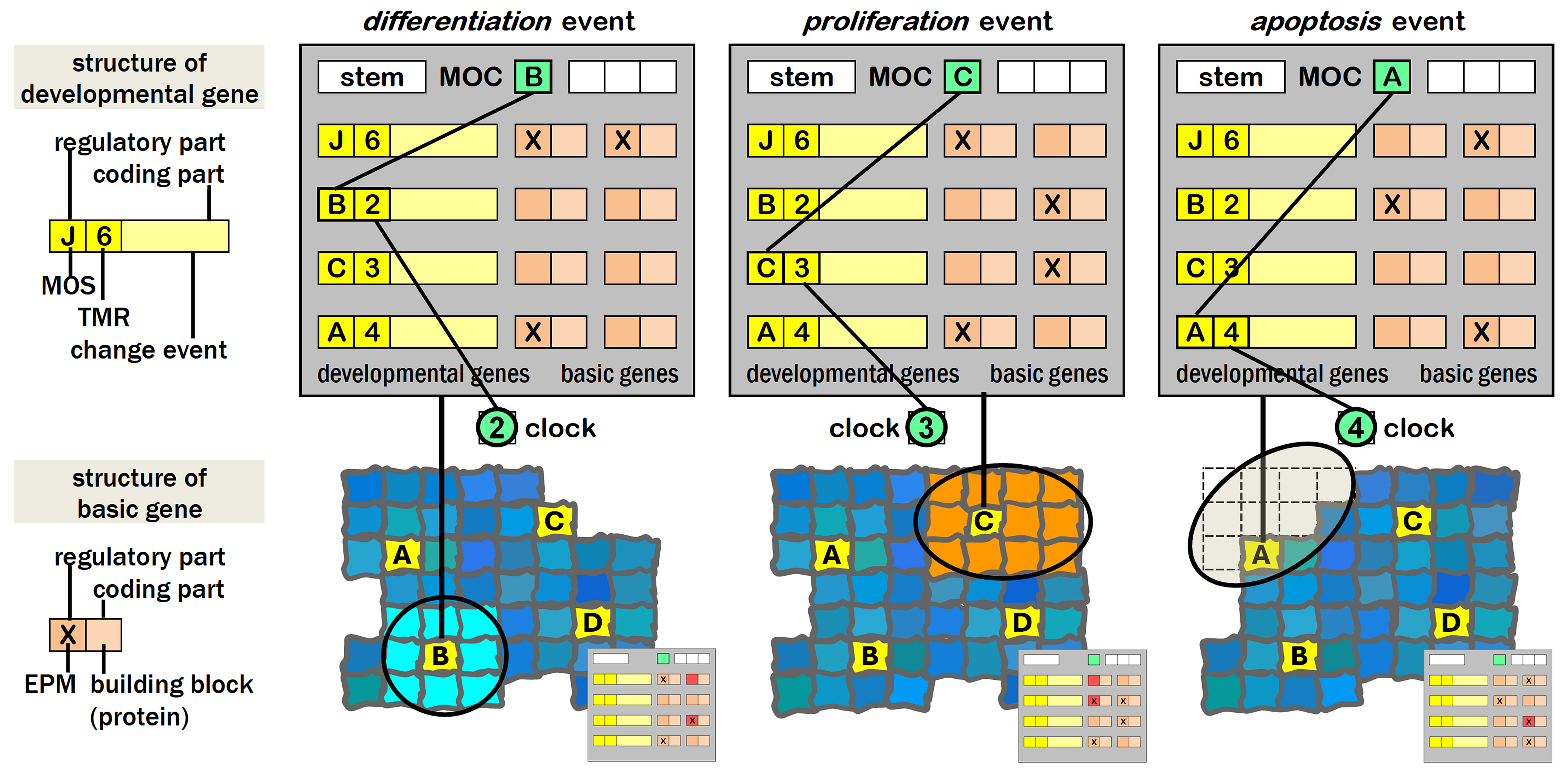}
\caption{
\normalsize
\if\afhead1 {\parok{figurex}} \fi
\if\afhead1 {\parok{caption}} \fi
The genome is divided into basic and developmental genes. Developmental genes encode change events that progressively reshape the epigenetic landscape of specific target cells. In this example, three developmental genes are sequentially activated in stem cells B, C, and A at clock values 2, 3, and 4. Each gene carries instructions defining the target cell population and the epigenetic modifications to be imposed on their initial state. These changes lead to the structural disabling or enabling of certain basic genes, as indicated in the bottom right corner of each panel. Depending on the encoded instructions, the resulting outcome may be a change in cellular behaviour, a proliferation event, or an apoptosis event. The sequence of change events shapes development.
}
\label{events}
\end{center} \end{figure*}

\if\afhead1 {\parok{phenotypes, cells, genome}} \fi
In the ET model, phenotypes emerge from cellular structures arranged on a grid and comprised of two cell types: \neudef{normal cells}, which form the majority, and \neudef{stem cells}, which are fewer and uniformly distributed. Acting as ``captains,'' stem cells orchestrate \neudef{change events} in synchrony with a global clock. Each cell’s genome consists of two gene classes: \neudef{developmental genes} governing developmental processes and \neudef{basic genes} responsible for all other cellular functions. Additionally, cell differentiation is defined by two non‑genetic elements: the \neudef{stem mark}, which distinguishes stem cells from normal cells, and the \neudef{master organisation code (MOC)}, which serves as the central regulator of differentiation.


\if\afhead1 {\parok{analogy for gene categories}} \fi
The category of \neudef{basic genes} largely overlaps with conventional structural genes, which encode proteins that form the organism’s essential components. In contrast, \neudef{developmental genes} contain the instructions guiding how these components are assembled into functional structures. Using a Lego analogy, basic genes correspond to the ``bricks'' (differing in shape, size, and color) while developmental genes act as the ``blueprint,'' providing exact assembly instructions (e.g., ``build two stacks of ten yellow bricks each, then connect them with two round black bricks to form a bridge''). Together, these gene classes coordinate to shape the organism’s form and function, though this analogy is highly simplified compared to biological complexity.

\color{colcg}
\if\afhead1 {\parok{gene structure}} \fi
Basic genes consist of a regulatory part and a coding part. The regulatory part can be flagged with an \neudef{epigenetic mark (EPM)}, which determines gene availability (genes marked with ``X'' are unavailable), while the coding part specifies a basic building block. Also developmental genes have a regulatory part and a coding part. The regulatory part comprises two elements: the \neudef{master organisation sequence (MOS)}, that can match the MOC; and the \neudef{timer (TMR)}, that can match the clock. When both elements are matched within a stem cell, the gene is activated and its coding part is executed. The coding part of a developmental gene encodes a \neudef{change event} (Fig.~\ref{events}) which switches the epigenetic marks of some basic genes in a set of cells called \neudef{change set}). This mechanism allows the orchestration of precise spatial and temporal genome activation during development.

\if\afhead1 {\parok{biological implement}} \fi
Biologically, the clock could be represented by a diffusible molecule propagated from a central source, while the MOC may correspond to master regulatory elements that initiate gene activation cascades within the cell. EPMs correspond to epigenetic modifications (e.g., histone or methylation marks) regulating gene accessibility. Developmental genes could involve long noncoding RNAs (lncRNAs), which are increasingly recognised as key regulators of chromatin structure and transcriptional control \cite{Rinn14}. Change events may correspond to writing, erasing and reading operations within the ``histone code'' framework \cite{Jenuwein01}. Modified epigenetic patterns can alter cellular behaviour, potentially triggering processes such as proliferation or apoptosis as specific outcomes. 

\if\afhead1 {\parok{generation of cell types}} \fi
A central feature of biological development is the generation of a wide spectrum of specialised cell types that collectively build the body’s organs and tissues. This transformation is governed by dynamic and highly regulated alterations in the epigenetic landscape, which modulate patterns of gene expression without changing the underlying DNA sequence. Epigenetic mechanisms act in concert to guide cells through successive stages of differentiation, gradually constraining their developmental potential while defining their specific identity \cite{Reik01, Jaenisch03}. The ET model captures these dynamics, offering a conceptual framework for understanding how developmental processes unfold across time and how cellular fate decisions are stabilised. Furthermore, the model can be combined with a \textit{genetic algorithm} that simulates biological evolution, giving rise to an evo-devo process able to generate complex structures \cite{Font10b}.
\color{black}

\begin{figure}[t] \begin{center} \hspace*{-0.00cm}
\includegraphics[width=16.00cm]{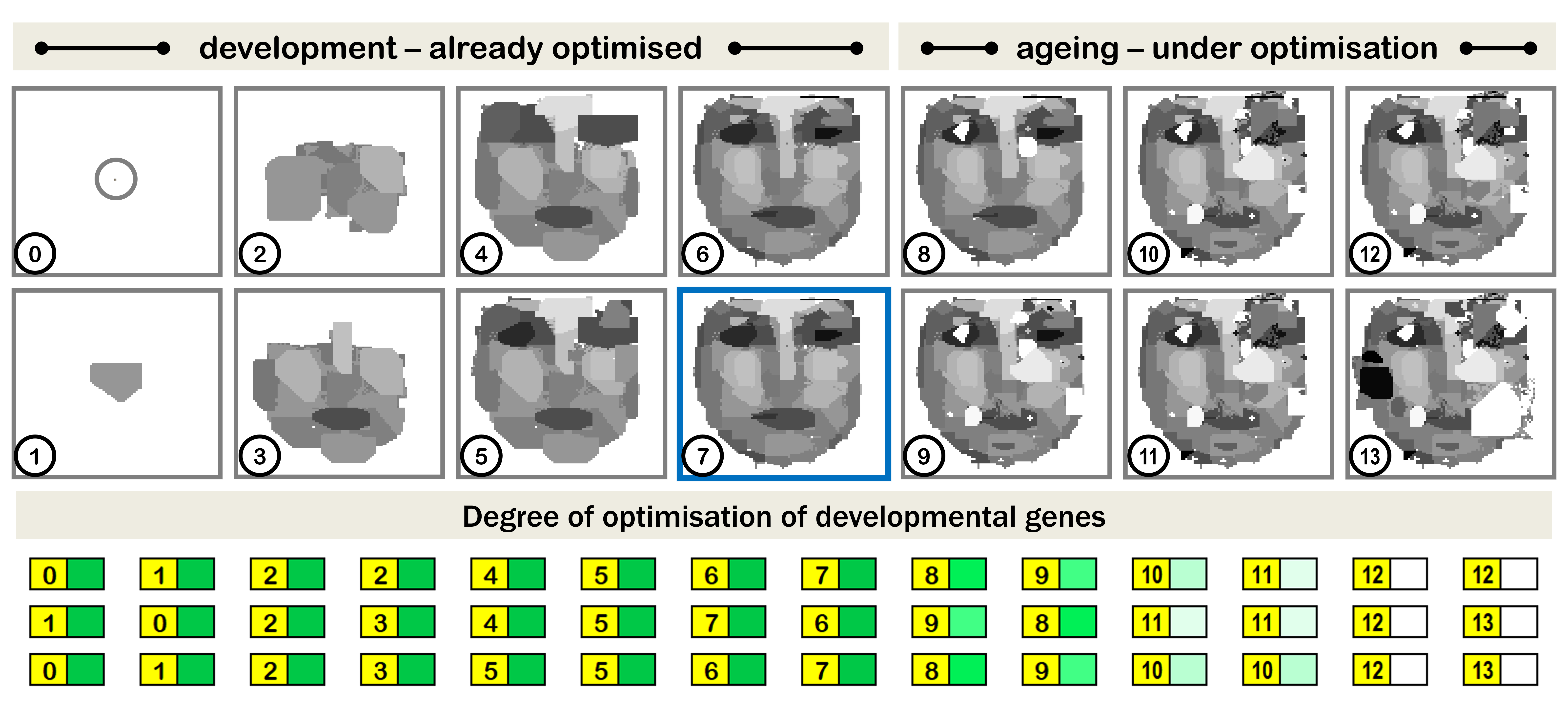}
\caption{
\normalsize
\if\afhead1 {\parop{figurex}} \fi
\if\afhead1 {\parop{caption}} \fi
\textbf{Upper panel.} Illustration of development and ageing for a 2-dimensional structure portraying a face. On the left the period of development: the structure grows from a single cell to the mature phenotype at the reproduction development step $ds = 7$ (blue frame), driven by optimised developmental genes. On the right the period of ageing: the quality of the structure deteriorates under the action of non-optimised developmental genes. \textbf{Lower panel.} The degree of optimisation of developmental genes (represented by the green shading) remains high until the point of reproduction, after which it gradually declines. Developmental genes with timer values $\leq 7$ display a high degree of optimisation, genes with timer values $\geq 8$ are less optimised, until their effects become completely pseudorandom.}
\label{pseudornd}
\end{center} \end{figure}

\color{colcg}
\if\afhead1 {\parok{fitness profile}} \fi
In the ET model, an individual progresses through N developmental steps, with fitness assessed across multiple stages following the designated \textit{reproduction step} r. The contribution of each step is weighted by coefficients that decrease with time yet sum to one, ensuring that earlier stages exert a stronger influence on overall fitness than later ones. This framework integrates the principle of kin selection \cite{West07}, here interpreted in a straightforward way: individuals are favoured if they maintain fitness beyond the point of reproduction, thereby enhancing the survival and success of their offspring. Biologically, this corresponds to the post-reproductive period during which parental care (or other forms of investment in kin) remains essential.
\color{black}

\if\afhead1 {\parok{pseudorandom gene expression}} \fi
Developmental genes programmed to activate at steps $>$ r, during what can be defined as the ageing phase, have a decreasing influence on the individual's fitness and are thus subject to weaker evolutionary optimisation. This leads to associated events exhibiting a ``pseudorandom'' nature: they appear random despite being encoded in the genome, and as such deterministic in nature. This pseudorandomness increases the likelihood that such events will have detrimental rather than beneficial effects on the individual. Under this framework, can be viewed as a \textit{continuation of development}, driven by non-optimised developmental genes that become active in specific stem cells after reproduction.

\if\afhead1 {\parok{painter analogy}} \fi
Using an analogy, the developmental process can be envisioned as a master robotic painter meticulously crafting a living masterpiece. Each brushstroke (defined by precise attributes such as timing, shape, colour, and texture) is encoded in the genome. Beginning with a blank canvas, the process gradually builds a complete artwork. Most theories of ageing implicitly assume that the painter halts its work once development is complete, leaving the painting vulnerable to passive decay—fading colours, cracking varnish, and the slow erosion of time. In contrast, the ET model posits that the painter never sets down the brush, but works seamlessly from conception until death. Yet as evolutionary pressure diminishes, the once-deliberate, optimised strokes grow increasingly erratic and disorganised, subtly distorting the painting over time. An illustration of this process is shown in Fig.~\ref{pseudornd}. 
 
\section{The ESTA hypothesis}
\label{sec:hypothesis}

\if\afhead1 {\parok{ESTA intro 1}} \fi
The model presented is consistent with classical evolutionary theories, which attribute ageing to a lack of evolutionary pressure. To uncover a deeper biological role for ageing, it is essential to acknowledge the extensive optimisation achieved through evolution, shaping highly refined biological designs across species. Given that evolution operates through random modifications, these changes are more likely to be harmful than beneficial for the individuals affected. Consequently, the effects of detrimental alterations should be observable in all evolving species. This raises a fundamental question: where do these effects become apparent? It seems there is a \textit{cause without an effect}. On the other hand, ageing appears to result from the accumulation of seemingly random, harmful events, yet the molecular mechanisms and the interaction between genetic and environmental factors remain elusive. Here, we encounter an \textit{effect without a cause}.

\if\afhead1 {\parok{ESTA intro 2}} \fi
This perspective presents an opportunity to unify two concepts: evolution as the cause of ageing and ageing as a consequence of evolution. More precisely, the proposition is that ageing primarily reflects the cumulative manifestation of genetically encoded epigenetic changes that are predominantly expressed during the post-reproductive phase. These late-acting modifications are not yet evolutionarily optimised but are instead subject to ongoing selection, functioning as somatic ``experiments'' through which evolution explores novel phenotypic variation. These experiments are often detrimental, leading to progressive physical decline and eventual death, while a small subset may produce beneficial adaptations, that evolution can exploit to shape future developmental trajectories. This hypothesis is referred to as the \neudef{Evolvable Soma Theory of Ageing (ESTA)}.  

\color{colcg}
\if\afhead1 {\parok{epigenetic process}} \fi
It is important to emphasise that this model posits these evolutionary experiments are orchestrated by genes that encode epigenetic modifications in specific target cells, activated at defined life stages. Evolution modifies these genes between generations, in the germline before conception, and no further mutations during life in somatic cells are needed in this model: epigenetic changes unfold as programmed. Epigenetic patterns in cells tend to lose coherence and become more randomised with age, in agreement with the proposed hypothesis. The concept of an \textit{epigenetic clock} \cite{Horvath13, Levine18}, defined by sets of DNA methylation markers that predict biological age across tissues and species, is consistent with ESTA if one assumes that the underlying process of epigenetic drift is genetically regulated. Nonetheless, the ESTA framework does not exclude contributions to ageing from other factors, such as somatic DNA mutations or environmental damage \cite{Woods25}.

\if\afhead1 {\parok{terminal addition}} \fi
The ESTA model is consistent with concepts such as ``terminal addition'' \cite{Gould77} and ``peramorphosis'' \cite{Alberch79}, which propose that evolutionary innovations are more likely to arise in late developmental stages. In this perspective, evolution proceeds in phases: early stages become progressively ``frozen'' and resistant to change, while later stages remain more flexible and open to experimentation (see Fig.~\ref{pseudornd}). This strategy permits the introduction of novel traits with minimal impact on fitness and reduced risk, while successful innovations can subsequently be shifted earlier in development through adjustments in their gene timers. The process is analogous to clinical trials, where new therapies are first tested in high-risk patients before being extended to healthier cohorts.
\color{black}

\if\afhead1 {\parop{assumptions}} \fi
As stated in \cite{Font25cs}, the ESTA hypothesis can be formalised through four key assumptions. First, development follows a programmed sequence unfolding from conception to death. This program encodes genetically encoded change events that modify the epigenetic landscape of target cells, fine-tuning the developmental process during the pre-reproductive period and shaping the body through a diverse range of cell types. Second, fitness is evaluated across multiple stages beginning at reproduction, accounting for the potential influence of kin selection. Third, the mutation rate is variable, with each genome base's mutation rate determined by the timer of the gene it belongs to: genes affecting later developmental stages experience higher mutation rates, effectively concentrating most mutations in the post-reproductive period. Finally, the timers of developmental genes are subject to evolutionary change, allowing events to be delayed or anticipated.

\color{colcg}
\if\afhead1 {\parok{ESTA vs SEAT/ past vs present evol}} \fi
ESTA can be contrasted with the Mutation Accumulation (MA), Antagonistic Pleiotropy (AP), and Disposable Soma (DS) theories, here collectively termed the Standard Evolutionary Ageing Theory (SEAT). Table~\ref{scentable} summarises their differences. In brief, SEAT views ageing as a byproduct of waning evolutionary engagement, whereas ESTA interprets ageing as evolution in action. Both share the core assumption that the fitness impact of genes declines with later expression, but they diverge in conclusion: SEAT argues that evolution acts mainly on early stages to \textit{maximise impact}, while ESTA posits that it acts on later stages to \textit{minimise risk}. Put differently, SEAT focuses on \textit{past} evolution, treating ageing as neglect, while ESTA highlights \textit{present} evolution, framing ageing as the outcome of ongoing, imperfect experiments required for innovation.

\if\afhead1 {\parok{ESTA vs SEAT/ basic vs developmental genes}} \fi
The concerns raised by AP and DS theories, where a mutation can be beneficial in one context but harmful in others, apply mainly to basic (protein coding) genes, which are broadly expressed across tissues and life stages. Changes in these genes might be advantageous under certain conditions but detrimental in most others and the net result is almost always negative. By contrast, developmental genes are subject to tighter spatial and temporal control, reducing the risk of harmful effects outside their intended context. With respect to MA, while selective pressure declines after reproduction, it does so gradually, and genes expressed in the post-reproduction period (late enough to reduce risk but still relevant to fitness) remain under evolutionary optimisation. Moreover, the evolvability of gene timers, a key feature of ESTA, provides additional flexibility by allowing the timing of gene activation to shift across generations.
\color{black}

\begin{table}[htbp] 
\centering
\small 
\begin{tabular}
{L{5.6cm}P{4.4cm}P{4.4cm}}
\toprule 
\textbf{THEORY} & \textbf{SEAT} & \textbf{ESTA} \\
\midrule 
\multicolumn{3}{l}
{\textit{\textbf{Ageing features}}} \\ 
\midrule
\textbf{Driver of development} & 
unknown (?) & 
epigenetic somatic changes \\
\textbf{Driver of ageing} & 
genetic somatic damages (?) & 
epigenetic somatic changes \\
\textbf{Indirect driver of ageing} & 
external or internal hits & 
genetic program \\
\textbf{Role of genetics in ageing} & 
maintenance and repair & 
active driver of ageing \\
\midrule
\multicolumn{3}{l}
{\textit{\textbf{Evolution features}}} \\ 
\midrule
\textbf{Fitness impact of genes} & 
declines with later expression & 
declines with later expression \\
\textbf{Mutation rate} & 
uniform across genome & 
dependent on gene timer \\
\textbf{Period under stronger selection} & 
before reproduction & 
after reproduction \\
\textbf{Developmental gene timers} & 
fixed (?) & 
evolvable \\
\bottomrule 
\end{tabular}
\caption{Comparison between SEAT and ESTA main features.}
\label{scentable}
\end{table}

\begin{figure*}[t] 
\begin{center} \hspace*{-0.5cm}
\includegraphics[width=15.0cm]{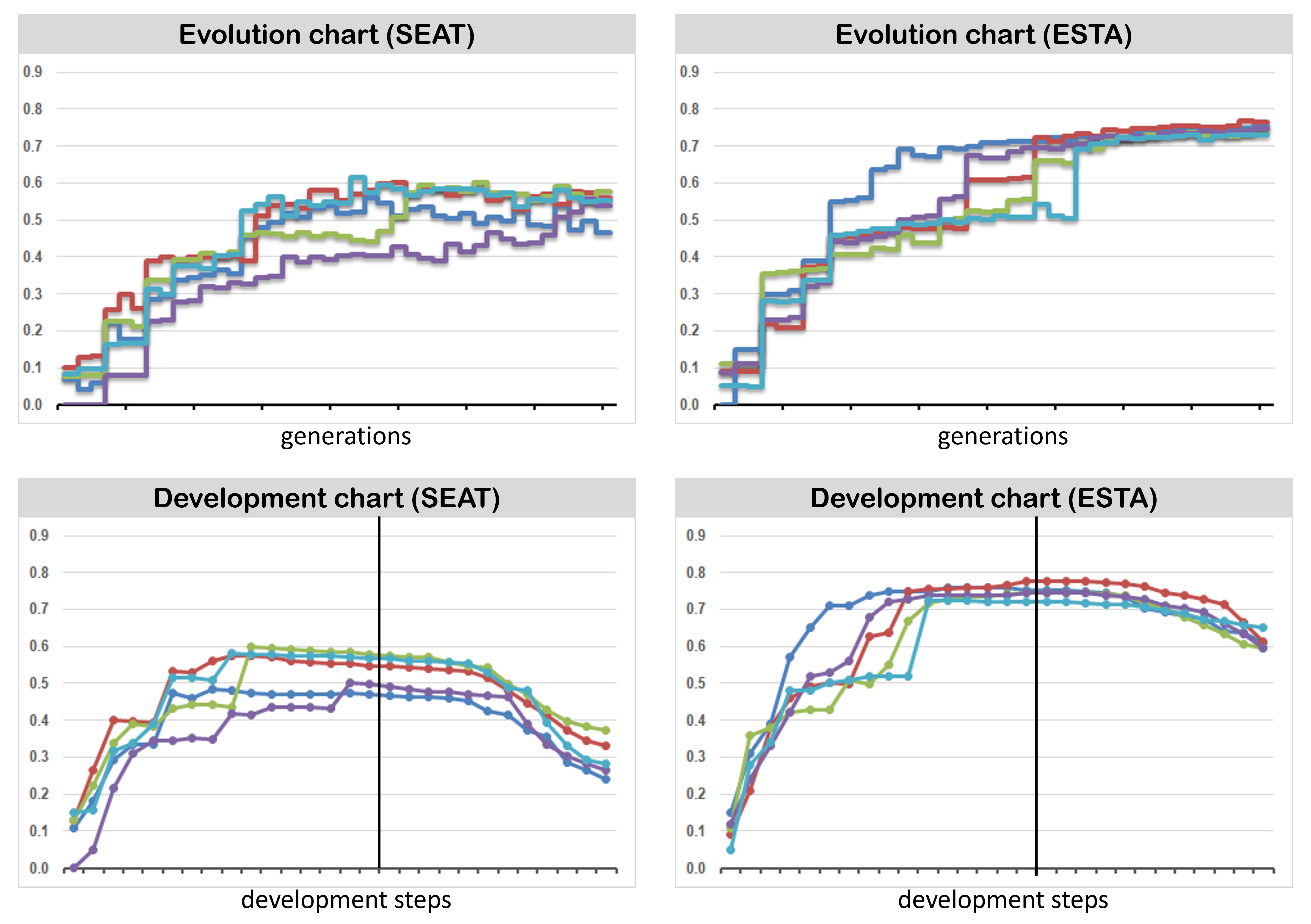}
\caption{
\if\afhead1 {\parok{caption}} \fi
\textbf{Evo-devo simulations for SEAT and ESTA.} For the two scenarios, the top panels display the mean reproduction fitness value of the population during evolution, for five independent evolutionary runs. The bottom panels show the mean fitness value of the population during development, captured in the last generation, for five independent evolutionary runs. For all panels, the y-axis represents the fitness value. In the top panels, the x-axis shows the generations, with each of the 8 ticks corresponding to 4000 generations, for a total of 32000 generations. In the bottom panels, the x-axis represents the development steps, ranging from 0 to 27, with a vertical line indicating the reproduction step at the end of the simulation. Further details can be found in \cite{Font25cs, Font25cs2}.}
\label{evodevocrt}
\end{center} 
\end{figure*}

\if\afhead1 {\parop{previous work, computer simul}} \fi
In summary, ESTA predicts a deep connection between evolution and ageing, across all evolving species. In previous work \cite{Font25cs, Font25cs2}, this connection was explored through computer simulations using the ET platform. The results demonstrated that ESTA achieves superior evolutionary-developmental (evo-devo) efficiency compared to SEAT (Fig.~\ref{evodevocrt}). Specifically, simulations conducted under the ESTA scenario showed faster fitness evolutionary progress and a more gradual fitness decline in development after reproduction. The results highlight that the core features of ESTA (in particular timer-dependent mutation rates and evolvable gene timers) are critical for its enhanced performance. In the remainder of this paper, we explore a significant implication of the proposed hypothesis for understanding human pathologies.


\section{Age-related diseases}
\label{sec:diseases}

\if\afhead1 {\parok{major diseases}} \fi
Cardiovascular diseases (CVD), cancer, and neurodegenerative disorders (including dementia) are among the leading causes of death globally, each representing a significant burden on public health. Cardiovascular diseases, including heart attacks and strokes, remain the top cause of mortality worldwide \cite{Lopez25}. Cancer, the second leading cause of death, encompasses a wide range of diseases characterised by uncontrolled cell growth, with lung, breast, and colorectal cancers being among the most prevalent \cite{Sung20}. Dementia, particularly Alzheimer's disease, is a growing concern as populations age, leading to progressive cognitive decline and loss of independence \cite{Livingston24}.

\if\afhead1 {\parok{metabolic syndrome}} \fi
Metabolic syndrome \cite{Saklayen18}, characterised by a cluster of conditions such as abdominal obesity, hypertension, dyslipidemia, and hyperglycemia, acts as a significant accelerator of the aforementioned diseases. It increases the risk of cardiovascular diseases by promoting atherosclerosis and hypertension \cite{Saklayen18}, exacerbates cancer progression through chronic inflammation and insulin resistance \cite{Braun19}, and is linked to cognitive decline and dementia, particularly vascular dementia, by impairing brain structure and function \cite{Machado22}. Together, cardiovascular diseases, cancer, dementia and metabolic syndrome are referred to by MD Peter Attia as the ``four horsemen of death'' \cite{Attia23}.

\if\afhead1 {\parok{risk factors, actionable}} \fi
Many risk factors contribute to the development of cardiovascular diseases, cancer, neurodegenerative disorders, and metabolic syndrome, often through complex interactions between genetic predispositions, environmental exposures, and lifestyle choices. Among the most prominent, poor diet and physical inactivity play a central role by promoting obesity, hypertension, dyslipidaemia, and insulin resistance, which are key drivers of CVD and metabolic disorders \cite{Saklayen18}. Chronic low-grade inflammation, frequently induced by smoking, air pollution, or excessive alcohol consumption, exacerbates these processes and has been strongly linked to endothelial dysfunction, carcinogenesis, and neurodegeneration \cite{Coussens02, Livingston24}.

\if\afhead1 {\parok{risk factors, actionable}} \fi
Psychological stress and sleep deprivation have also emerged as critical contributors, influencing metabolic pathways and immune responses that accelerate disease onset \cite{Kivimaki23}. Moreover, socioeconomic disparities exacerbate exposure to these risks, with lower-income populations experiencing higher rates of morbidity due to limited access to healthcare, nutritious food, and safe environments for physical activity \cite{Braveman14}. Collectively, these modifiable factors are referred to as the ``exposome,'' a concept that encompasses the totality of environmental exposures experienced by an individual across their lifetime, and their cumulative impact on health and disease risk \cite{Woods25}.

\if\afhead1 {\parok{risk factors, genetic}} \fi
Genetic predisposition plays a significant role in the development of the aforementioned diseases. Genome-wide association studies (GWAS) have identified multiple loci associated with an increased risk of cardiovascular events, particularly variants in the 9p21 region, which influence atherosclerosis and coronary artery disease susceptibility independent of traditional risk factors \cite{Schunkert11}. Similarly, inherited mutations in oncogenes (BRCA1/2, KRAS) and tumor suppressor genes (TP53, RB1) contribute to cancer susceptibility by altering cellular proliferation and DNA repair mechanisms \cite{Vogelstein13}. Moreover, low-effect genetic variants interact with environmental factors to shape cancer risk, reflecting the complexity of gene–environment interplay.

\if\afhead1 {\parok{risk factors, genetic}} \fi
In dementia, genetic variations in the APOE gene, particularly the APOE4 allele, are strongly linked to an elevated risk of Alzheimer's disease through mechanisms involving amyloid-beta aggregation and impaired lipid metabolism \cite{Jansen19}. Additional risk loci identified by genome-wide association studies suggest a broader polygenic basis, implicating genes involved in synaptic function, endocytosis, and neuroinflammation. Metabolic syndrome also has a hereditary component, with polymorphisms in genes related to insulin signaling (TCF7L2), lipid metabolism (FTO), and inflammation contributing to obesity, dyslipidaemia, and type 2 diabetes \cite{Mahajan18}. Despite the influence of genetic factors, gene-environment interactions remain crucial, as lifestyle modifications can mitigate disease expression even in high-risk individuals. 

\if\afhead1 {\parok{risk factors, age}} \fi
However, the single most significant risk factor for most diseases is age. For cardiovascular diseases, the risk increases exponentially with age, with individuals aged 60 and older accounting for over 75\% of CVD-related deaths globally \cite{North12}. Similarly, cancer incidence rises sharply with age, with approximately 88\% of all cancer cases diagnosed in individuals aged 50 and older, and the median age of cancer diagnosis being 66 years \cite{Sung20}. Dementia prevalence also escalates with age, doubling every 5 years after the age of 65, and affecting over 30\% of individuals aged 85 and older \cite{Livingston24}. Metabolic syndrome shows a similar age-dependent trend, with prevalence increasing from 20\% in individuals aged 40-49 to over 40\% in those aged 60 and older \cite{Saklayen18}.

\section{The mother of age-related diseases}

\if\afhead1 {\parok{link evolution-ageing}} \fi
As described in section \ref{sec:hypothesis}, within the ESTA framework ageing reflects the accumulation of genetically encoded epigenetic modifications, expressed during the post-reproductive phase. These late-acting changes are not yet fully optimised by evolution; instead, they represent ongoing somatic ``experiments'' that allow evolution to explore novel phenotypic variation. Unlike SEAT, which frames ageing as a consequence of limited evolutionary pressure, ESTA fundamentally redefines ageing as evolution in action. At its core, ESTA establishes a causal link between evolution and ageing, proposing that ageing is caused by evolution itself.

\if\afhead1 {\parok{link ageing-diseases}} \fi
In Section \ref{sec:diseases}, we discussed the well-established association between ageing and a broad spectrum of diseases whose incidence increases markedly with advancing age and ultimately contributes to mortality. Prominent examples include cardiovascular diseases, cancer, neurodegenerative disorders, and metabolic syndrome, all of which display a pronounced age-dependent rise in incidence. We propose that this relationship is not merely correlative but fundamentally causative: age-related diseases arise (primarily) from genetically programmed epigenetic modifications that alter key physiological pathways during the ageing process. In this perspective, the distinction between ``normal'' or ``healthy'' ageing and age-related pathologies is primarily quantitative rather than qualitative, reflecting a continuum of effects mediated by the same underlying biological mechanisms.

\if\afhead1 {\parok{putting two links together}} \fi
By combining the two premises---that ageing is driven by evolution and that diseases arise as a consequence of ageing---we arrive at a striking conclusion: evolution itself is the root cause of age-related diseases. This perspective reframes our understanding of pathology, positioning evolution not only as the force that enables species to survive and thrive but also as the underlying driver of biological deterioration. In this sense, the very mechanism that ensures the survival of a species also imposes an inevitable cost---one that manifests as progressive decline, disease, and mortality. Once again, it is important to emphasise that this is not a matter of evolution ``turning a blind eye'': rather, it reflects an active, genetically encoded process.

\if\afhead1 {\parok{diseases as system attractors}} \fi
If age-related diseases are viewed as byproducts of evolutionary experimentation on epigenetic pathways, a pertinent question arises: why do these processes specifically result in the diseases commonly encountered in clinical practice? One plausible explanation is that these diseases are the most likely outcomes when the biological system is subjected to random perturbations around the optimal equilibrium reached at the time of sexual maturation. In dynamical systems' theory, \textit{an attractor} is a state that a system tends to evolve toward over time, regardless of its initial starting conditions \cite{Strogatz15}. In this context, age-related diseases may act as attractors within the system under conditions of random disturbance.

\if\afhead1 {\parok{no patterns of gene mutations found in cancer}} \fi
Despite significant advancements in cancer genomics, identifying consistent patterns of genetic mutations associated with cancers remains a challenge. While certain mutations, such as those in TP53, BRCA1, and KRAS, are frequently encountered in many cancer types, the overall mutational landscape is highly heterogeneous \cite{Martincorena15}. While large-scale sequencing efforts have identified recurrent mutations in oncogenes and tumor suppressors, the presence of extensive genetic heterogeneity suggests that mutational patterns are unlikely to fully account for cancer development across different individuals and tissue types \cite{Salk10}. A possible explanation is that most common cancers may have an inherently epigenetic origin \cite{Baylin16}, in accordance with our hypothesis.

\if\afhead1 {\parok{no patterns of gene mutations found in many other diseases}} \fi
Other age-associated diseases such as dementia and cardiovascular disease have not yielded consistent patterns of gene mutations in genome-wide association studies (GWAS). While GWAS have successfully identified numerous risk loci for late-onset diseases, these associations often account for only a fraction of heritability and rarely involve recurrent mutations or high-penetrance variants \cite{Keller12}. For example, in Alzheimer's disease, aside from the well-established \textit{APOE} $\varepsilon4\varepsilon4$ allele, most identified variants exhibit modest effect sizes and are dispersed across diverse pathways \cite{Lambert13}. The absence of strong mutational patterns in such diseases, for which ageing itself represents the primary risk factor, suggests that these conditions may arise more from dysregulated epigenetic or systemic processes than from fixed genetic defects \cite{laTorre23}.

\if\afhead1 {\parok{exception: infectious disease}} \fi
Age-related diseases account for a large portion of human pathology. Among non age-associated diseases, two categories are particularly noteworthy. The first category is represented by infectious diseases, which can affect individuals across all age groups, including both the young and the elderly, though their impact can vary significantly between these groups. For older adults, especially those with weakened immune systems or underlying chronic conditions, infectious diseases can also be particularly dangerous, often leading to complications such as sepsis or organ failure \cite{Feng20}. Infectious diseases are excluded from the set of conditions potentially driven by programmed epigenetic changes, as their onset is primarily determined by external pathogenic agents rather than internal biological processes.

\if\afhead1 {\parok{exception: single-gene diseases}} \fi
The other category is represented by monogenic diseases, caused by mutations in single (usually protein-coding) genes. These conditions, (including disorders such as e.g. cystic fibrosis, sickle cell anaemia, Huntington’s disease, and Tay-Sachs disease) are generally rare, often manifesting with severe clinical phenotypes and an early age of onset. For example, cystic fibrosis, caused by mutations in the CFTR gene, affects approximately 1 in 2500 newborns in populations of European descent \cite{Riordan89}. Haemophilia, an X-linked disorder usually caused by mutations in the F8 or F9 genes, impairs the blood's ability to clot and affects approximately 1 in 5000 male births \cite{White01}. Also monogenic diseases can be excluded from the category of conditions driven by programmed epigenetic changes. Using the painting analogy introduced in Section \ref{sec:model}, they can be likened to defects in the base colours used by the painter: fundamental flaws in the raw materials rather than insufficient optimisation of the developmental program.

\if\afhead1 {\parok{other factors}} \fi
In this section, we have conjectured that genetically encoded epigenetic changes represent the principal driver of ageing and age-related diseases. This, however, does not exclude the contribution of additional factors. Unhealthy eating habits, physical inactivity, and somatic mutations induced by exposure to radiation or chemical substances may exert effects superimposed on the epigenetic program, thereby accelerating the process of damage accumulation. Notably, evidence indicates that disease incidence in identical twins is similar but not identical. This has been observed in cardiovascular disease \cite{Song15}, cancer \cite{Mucci16}, neurodegenerative disorders \cite{Gatz06}, and metabolic syndrome \cite{Poulsen09}.

\section{Conclusions}

\if\afhead1 {\parok{recap}} \fi
This work explored some implications of the ESTA model, which frames development and ageing as a continuous process driven by genetically encoded epigenetic changes in specific cellular sets. Within the ESTA framework, ageing reflects the accumulation of these epigenetic modifications, primarily expressed during the post-reproductive phase. These late-acting changes are not yet fully optimised by evolution; instead, they represent ongoing somatic ``experiments'' that allow evolution to explore novel phenotypic variation. Consequently, ageing can be viewed as evolution in real-time.

\if\afhead1 {\parok{conclusions}} \fi
At the same time, old age is the strongest risk factor for major diseases, including cardiovascular diseases, cancer, neurodegerative disorders, and metabolic syndrome. We argue that this relationship is not merely correlational but causal: the same processes that drive ageing also underlie age-associated diseases. In this view, evolution is not only responsible for ageing but also for the emergence of diseases that accompany it, making it the ultimate origin of most age-related pathologies. This trade-off highlights the dual nature of evolution: a force of adaptation and innovation, yet also the architect of biological fragility.

\if\afhead1 {\parok{future work}} \fi
Like SEAT, ESTA acknowledges that ageing occurs in a period of reduced evolutionary pressure. However, whereas SEAT suggests that this leads to the accumulation of deleterious alleles, ESTA argues that such alleles are not stable but highly variable, allowing evolution to ``learn'' from their phenotypic effects. In other words, these alleles are not merely evolutionary leftovers but represent a dynamic exploration of novel phenotypic space driven by genetic variation. This prediction is potentially susceptible of being tested through evolutionary genetics experiments using animal models such as \textit{Drosophila Melanogaster} \cite{Frankel21}. This will be matter for future work.

\bibliographystyle{apalike} 
\bibliography{lxevoldisc}
 
\end{document}